# Direct Visualization of Room-temperature Stair-stepped Quantum Spin Hall States in $Bi_4Br_4$

Zhiqiang Hu[1,2,*], Yuqi Zhang[3,*], Yuyang Wang[1,2], Kebin Xiao[1,2], Xiang Li[3], Zhiwei Wang[3,4,5], Huaixin Yang[6], Yugui Yao[3,4,5,†], Qi-Kun Xue[1,2,7,8,9,†] and Wei Li[1,2,9,†]

[1]*State Key Laboratory of Low-Dimensional Quantum Physics, Department of Physics, Tsinghua University, Beijing 100084, China*
[2]*Frontier Science Center for Quantum Information, Beijing 100084, China*
[3]*Centre for Quantum Physics, Key Laboratory of Advanced Optoelectronic Quantum Architecture and Measurement (MOE), School of Physics, Beijing Institute of Technology, Beijing 100081, China*
[4]*Beijing Key Lab of Nanophotonics and Ultrafine Optoelectronic Systems, Beijing Institute of Technology, Beijing 100081, China*
[5]*International Center for Quantum Materials, Beijing Institute of Technology, Zhuhai 519000, China*
[6]*Beijing National Laboratory for Condensed Matter Physics, Institute of Physics, Chinese Academy of Sciences, Beijing 100190, China*
[7]*Beijing Academy of Quantum Information Sciences, Beijing 100193, China*
[8]*Southern University of Science and Technology, Shenzhen 518055, China*
[9]*Hefei National Laboratory, Hefei 230088, China*

∗These authors contributed equally to this work.
†Corresponding authors. E-mails: ygyao@bit.edu.cn; qkxue@mail.tsinghua.edu.cn; weili83@tsinghua.edu.cn;

## Abstract
Topological insulators host exotic quantum phenomena such as the quantum spin Hall (QSH) effect, which enables dissipationless one-dimensional edge conduction. Realizing such states at room temperature and on a macroscopic scale is essential for energy-efficient electronics and quantum technologies, yet remains a fundamental challenge due to material limitations. Here, using microwave impedance microscopy, we directly visualize robust QSH states persisting up to 300 K in $α$-Bi$_4$Br$_4$ nanowires. This stability and scalability are enabled by a "stair-stepped" stacking configuration, a multilayer geometry in which QSH edge states from individual layers remain spatially decoupled. This configuration circumvents the stringent alignment and layer-number constraints of previous proposals, allowing robust stair-stepped QSH (SS-QSH) conduction in structures several micrometers long and hundreds of nanometers high. Magnetic-field and temperature-dependent measurements confirm their intrinsic topological nature. Crucially, the SS-QSH and bulk signals scale with nanowire height, verifying the stair-stepped origin. Our results are also successfully reproduced by finite-element analysis simulations. This work establishes $α$-Bi$_4$Br$_4$ as a practical platform for high-temperature topological electronics and demonstrates a generalizable stacking strategy for designing scalable QSH systems.

**Keywords**: $α$-Bi$_4$Br$_4$ nanowires, room-temperature quantum spin Hall states, microwave impedance microscopy, stair-stepped stacking, magnetic and temperature dependence

## INTRODUCTION

Two-dimensional (2D) topological insulators possess a pair of counterpropagating helical QSH edge states surrounding the insulating bulk interior [1-3]. From the band-structure perspective, these time-reversal-symmetry protected QSH states arise from spin-momentum-locked one-dimensional (1D) Dirac bands residing within the 2D bulk band gap, thus enabling spin-resolved dissipationless edge transport [4-6]. Since the initial discovery of QSH states [7-9], substantial efforts have been devoted toward achieving robust room-temperature QSH systems with larger size [10-18], which are vital for applications such as spintronic devices and low-dissipation quantum circuits [12,19-22]. However, although significant progress has been made in certain aspects [13,14,23], the realization of practical QSH-based devices remains challenging, mainly constrained by material choice, small size, etc.

$α$-Bi$_4$Br$_4$ is a recognized platform for topological states [24-38] and holds great potential for realizing high-temperature QSH states owing to its large band gap exceeding 200 meV [24,25,30,32,33,36]. As illustrated in Fig. 1a, the crystal structure of layered $\alpha$-Bi$_4$Br$_4$ consists of quasi-one-dimensional chains extending along the *b*-axis direction (see also Fig. S1). Both interlayer and intralayer interactions are governed by van der Waals forces, with adjacent layers rotated by 180°, giving rise to its distinctive topological properties. Theoretically, monolayer Bi$_4$Br$_4$ has been predicted to be a QSH insulator [24], and QSH states have been expected to persist at monolayer steps on bulk Bi$_4$Br$_4$ substrates [25]. Scanning tunneling microscopy (STM) measurements have revealed an enhanced local density of states near the edges of monolayer Bi$_4$Br$_4$, consistent with the presence of QSH states [32]. However, STM probes local electronic density rather than direct conduction. The unambiguous confirmation of QSH physics, especially in multilayer, device-relevant architectures, requires direct visualization of dissipationless edge conduction.

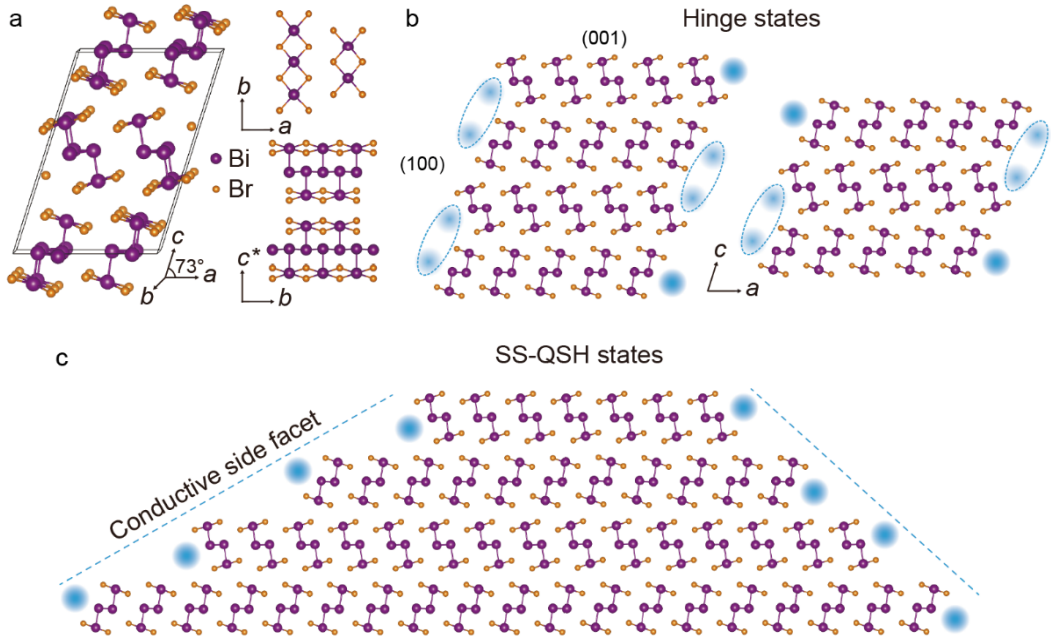

**Figure 1.** Stair-stepped stacking $α$-Bi$_4$Br$_4$. (a) Crystal structure of $α$-Bi$_4$Br$_4$. The angle between $a$- and $c$-axes is 73°. (b) Schematics of higher-order hinge states in $ac$ plane. QSH states in each layer are marked by blue dots. Layer edges are strictly aligned along the $c$-axis and QSH states from adjacent layers gap out in pairs (blue dashed circles). Left and right panels: configuration of hinge states with even and odd layer number. (c) Schematics of SS-QSH states in $ac$ plane. Spatially separated monolayer-step edges stack vertically and decoupled QSH states from each layer forms a conductive side facet with much shallower inclination angle than 73°.

For practical QSH-based devices, larger-scale systems with multiple Bi$_4$Br$_4$ layers are desirable. The transition from 2D to 3D, however, critically depends on the stacking geometry. When layer edges are strictly aligned along the *c*-axis, corresponding to a 73° inclination between *a*- and *c*-axes (Fig. 1b), QSH states from adjacent layers can hybridize strongly, gapping out in pairs [28,32,37]. The system then enters into a higher-order topological insulator (HOTI) phase, retaining only a single pair of QSH states localized at the intersection of (001) and (100) surfaces, known as hinge states [26,28,29,32,34,36,37]. These hinge states reside either on the same side (even layer number) or opposite sides (odd layer number) of the structure (left and right panels in Fig. 1b) [28,32,37]. While the HOTI phase has been confirmed in several studies, all demonstrations have relied on cleaved samples comprising only a few layers [29,32,36,37]. Strict edge alignment along the *c*-axis is relatively achievable in such few-layer systems but becomes exceedingly difficult to maintain in larger structures, rendering hinge states inherently unstable. Furthermore, determining the exact layer count in large systems is challenging, adding uncertainty to the predicted locations of hinge states and thereby limiting their practical utility.

A simpler stacking configuration featuring stair-stepped edges is presented in Fig. 1c [25]. This structure comprises numerous vertically stacked monolayer steps. The QSH states localized at each step edge are spatially separated and thus remain decoupled, collectively forming a conductive side facet with a much shallower inclination angle than 73° [30,35]. Consequently, the behavior of these SS-QSH states is largely independent of the precise layer count in larger systems. Given that edge states in monolayer steps can persist up to room temperature [32], SS-QSH states are also anticipated to maintain stability under ambient conditions. Owing to these advantages, SS-QSH states represent a highly promising configuration for practical applications.

In this work, we employ microwave impedance microscopy (MIM) to directly resolve the local conductivity of SS-QSH states in micron-scale $\alpha$-Bi$_4$Br$_4$ nanowires synthesized via physical vapor deposition (see also Section 1 in Supplementary Material). Unlike tunneling-based techniques that probe density of states of electrons, MIM provides unambiguous conductance mapping, enabling the real-space imaging of dissipationless conductive edge at the mesoscopic scale. Our results demonstrate that the stair-stepped stacking configuration successfully hosts robust SS-QSH states at room temperature, fulfilling its theoretical promise without stringent constraints on layer alignment or count. The following sections detail our imaging of these states, their magnetic-field and temperature resilience, and the height-dependent signatures that conclusively distinguish the stair-stepped origin.

## RESULTS
### Room-temperature SS-QSH states in $\alpha$-Bi$_4$Br$_4$ nanowires

MIM is a scanning probe technique well-suited for investigating topological materials, offering distinct advantages such as non-contact operation and a configuration that requires no fabricated electrodes on the sample [39-50]. The schematic of device structure and MIM setup is presented in Fig. 2a. The technique outputs two channels, including real and imaginary channels ($d$(MIM-Re)/$dz$ and $d$(MIM-Im)/$dz$, or directly called MIM-Re and MIM-Im in this work as explained in Section 2 in Supplementary Material), which carry information about the complex tip-sample impedance. In $\alpha$-Bi$_4$Br$_4$ nanowire system, the conductivity contrast between the SS-QSH and bulk states is the primary source of signal variations in both MIM channels. Simulated microwave response curves as a function of resistivity, obtained from finite-element analysis (FEA), are presented in Fig. 2b (see more details in Figs S3a-S3c). The MIM-Im signal varies monotonically with resistivity, allowing for direct qualitative comparison of conductivity. In contrast, the MIM-Re channel exhibits a symmetric lineshape with a central peak; it is typically used in conjunction with MIM-Im to constrain the resistivity range of the measured states.

Figure 2c shows the typical AFM topography of a nanowire. The inclination angles of its two side facets measure 25.0° and 28.4° (Fig. 2d), both far smaller than the 73° angle between *a*- and *c*-axes, in agreement with the stair-stepped model. MIM-Im images acquired at 4.2 K and 300 K using tapping and contact modes are presented in Figs 2e and 2f. Conductive channels along both side facets are clearly resolved and identified as SS-QSH states (two high peaks in line profile in Fig. 2e). A slight signal intensity difference between the two states stems from their differing facet angles. The circular textures visible in Figs 2c and 2e are surface contaminants introduced during sample transfer, which can be readily removed by contact-mode scanning (Fig. 2f); the unaffected SS-QSH states attest to their topological robustness. Note that Figs 2e and 2f were acquired on two different nanowires using tapping and contact modes, respectively, which accounts for the apparent differences in the edge-state signals. The MIM measurement process, peak-width analysis, and a detailed comparison between contact and tapping modes are provided in Figs S2 and S4. All subsequent MIM measurements used tapping mode for tip protection. These SS-QSH states remain clearly resolvable at 300 K (Figs 2f and S4), demonstrating exceptional thermal stability enabled by the wide band gap of α-Bi₄Br₄. They fully encircle a 16-μm-long nanosheet (Fig. S5) and appear on all three facets of a 450-nm-high nanowire (Fig. S6), confirming their compatibility with large-scale systems.

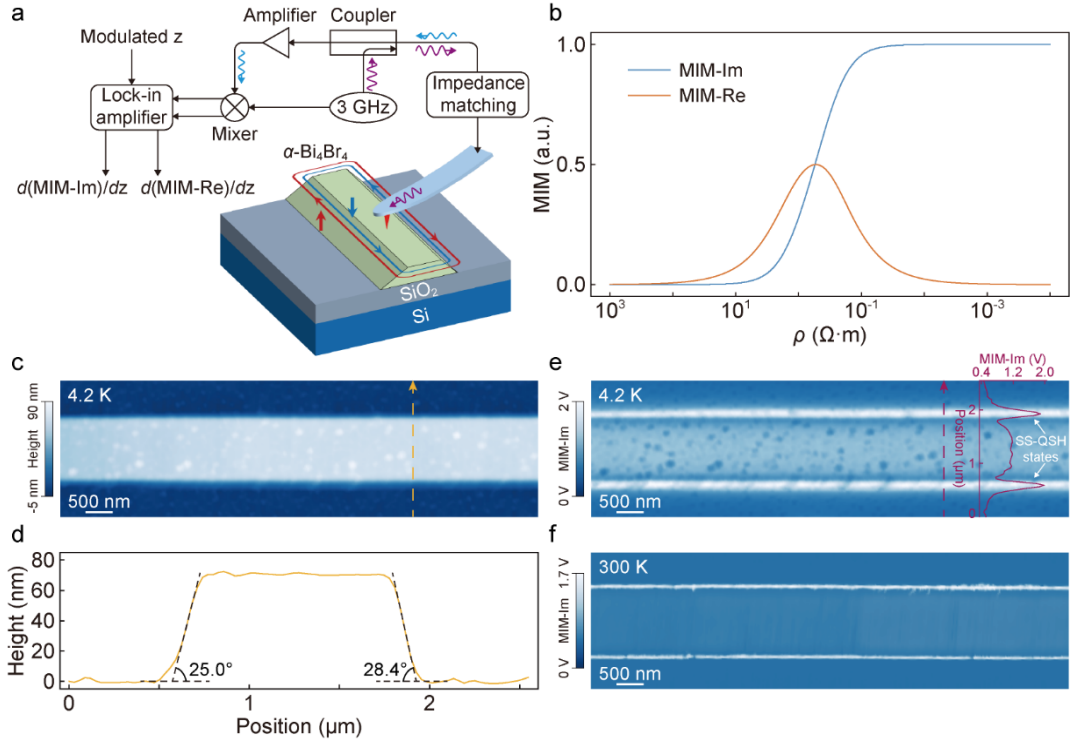

**Figure 2.** Visualizing SS-QSH states in α-Bi₄Br₄ nanowires. (a) Schematic of device structure and MIM setup. A pair of QSH states (spin-up in red, spin-down in blue) encircling α-Bi₄Br₄ nanowire is depicted as representative. (b) Simulated MIM-Im and MIM-Re signals as a function of resistivity (MIM response curves) using FEA. (c) AFM topography of a nanowire. (d) Height profile along the yellow dashed line in (c). The inclination angles of left- and right-side facets are 25.0° and 28.4°, respectively. Note that the position- and height-axes are scaled differently. (e, f) MIM-Im images of two nanowires measured at 4.2 K and 300 K, respectively. The purple profile in (e) is obtained along the purple dashed arrow. (e) is obtained simultaneously with (c) under tapping mode, while (f) is obtained under contact-mode.

## Temperature dependence of SS-QSH and bulk states

A detailed investigation was conducted on the temperature dependence of both SS-QSH and bulk states in the nanowires. MIM-Im and MIM-Re images of a nanowire were simultaneously acquired at 77 K (Figs 3a and 3b) and line profiles are shown in Fig. 3c. Both Single- and Average-profiles were taken for universality of results. The left and right peaks (dips) in Im (Re)-channels correspond to SS-QSH states of the nanowire. Careful MIM phase calibration and temperature modification acquired from standard sample are presented in Fig. S7. Complete data are presented in Fig. S8. Due to temperature-related fluctuations in the microwave circuit, the signal-to-noise ratio of MIM-Re signals is considerably lower than that of MIM-Im signals, limiting its use primarily to auxiliary interpretation. Obviously, SS-QSH states exhibit much higher MIM-Im signals than the relatively insulating bulk (upper panel of Fig. 3c). As for MIM-Re signals, SS-QSH states are located farther from the MIM-Re symmetry axis compared with bulk states (Fig. 3f), resulting in lower MIM-Re signals. Therefore, SS-QSH states manifest as two dips in Re-channel on two side facets of the nanowire (lower panel of Fig. 3c).

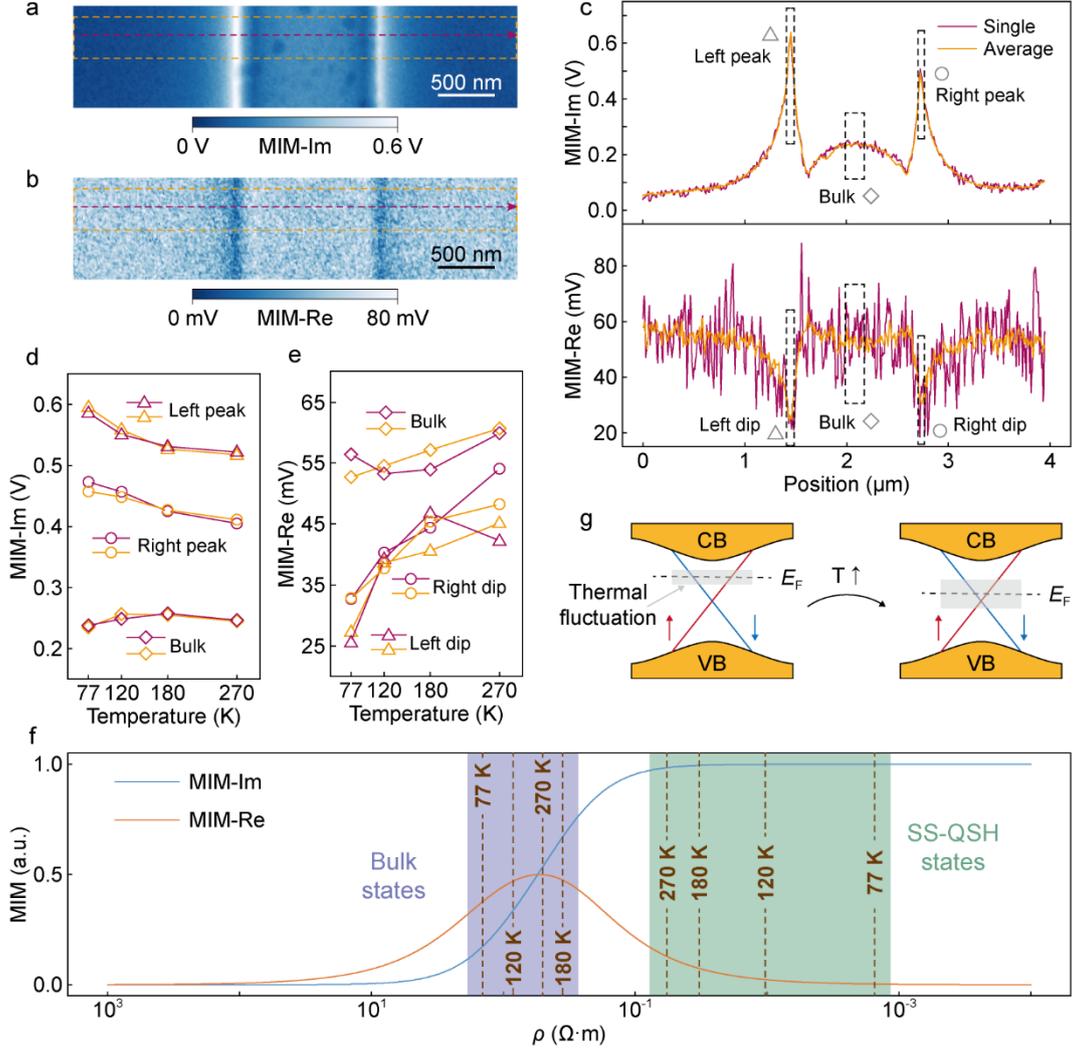

**Figure 3.** Temperature dependence of SS-QSH and bulk states. (a, b) MIM-Im and MIM-Re images of a nanowire measured at 77 K. (c) MIM-Im and MIM-Re line profiles from (a) and (b). Single-profiles are obtained along the purple dashed arrows and Average-profiles are obtained by averaging the yellow dashed rectangle area. (d, e) Temperature response of SS-QSH and bulk states. The values of peaks and dips are extracted by averaging their corresponding gray dashed rectangles in (c). Single- and Average-profile data are marked with purple and yellow, respectively. (f) Estimated resistivity of SS-QSH and bulk states in simulated response curves. The light green and purple areas represent their possible ranges and are intentionally enlarged for clarity. (g) Schematics illustrating temperature effects on SS-QSH states. Dirac-cone QSH states are represented by spin-up (red) and spin-down (blue) bands. Gray shaded areas depict temperature-induced thermal fluctuations. CB (conduction band), VB (valence band), and $E_F$ (Fermi level) are labeled.

Temperature-dependent behaviors of SS-QSH and bulk states are summarized in Figs 3d and 3e. We use averaged SS-QSH and bulk values to reveal their variations (gray dashed rectangles in Fig. 3c). For SS-QSH states, as the temperature rises from 77 K to 270 K, MIM-Im signals decrease monotonically while retaining relatively high values, indicating diminished yet significant conductivity. This trend suggests that SS-QSH states are only moderately suppressed by temperature, likely owing to the large band gap of α-Bi$_4$Br$_4$ [24,25,30,32,33,36]. Meanwhile, MIM-Re signals of SS-QSH states roughly increase monotonically with temperature. Based on the temperature evolution of MIM-Im and MIM-Re signals, SS-QSH states are confirmed to reside in the highly conductive region of the simulated response curves (light green-colored region in Fig. 3f). For bulk states, MIM-Im signals increase between 77 K and 180 K, followed by a slight decrease at 270 K, indicating that the conductivity first enhances and then diminishes upon heating. The initial enhancement is consistent with the intrinsic semiconducting characteristic of α-Bi$_4$Br$_4$. In contrast, the subsequent diminishment may be associated with the emergence of a local resistivity maximum between 180 K and 270 K, which has been reported by previous studies and is attributed to a Lifshitz transition induced by the downward shift of the Fermi level [31,33]. In addition, MIM-Re signals of bulk states also roughly increase monotonically with temperature, suggesting that bulk states lie in the intermediate region of the simulated response curve (light purple-colored region in Fig. 3f).

A possible explanation illustrating temperature effects on SS-QSH states is displayed in Fig. 3g. With increasing temperature, two main effects are prominent: a gradual downward shift of the Fermi level [31,33] and an enhancement of thermal fluctuations. These effects allow a small fraction of electrons from the lower half of the Dirac cone to participate in the conduction of SS-QSH states. This participation enhances spin-flip process, where electrons from spin-up (spin-down) channel are partially converted into counterpropagating spin-down (spin-up) electrons, leading to a reduced conductivity. However, due to the small proportion of converted electrons, SS-QSH states remain strong at 300 K.

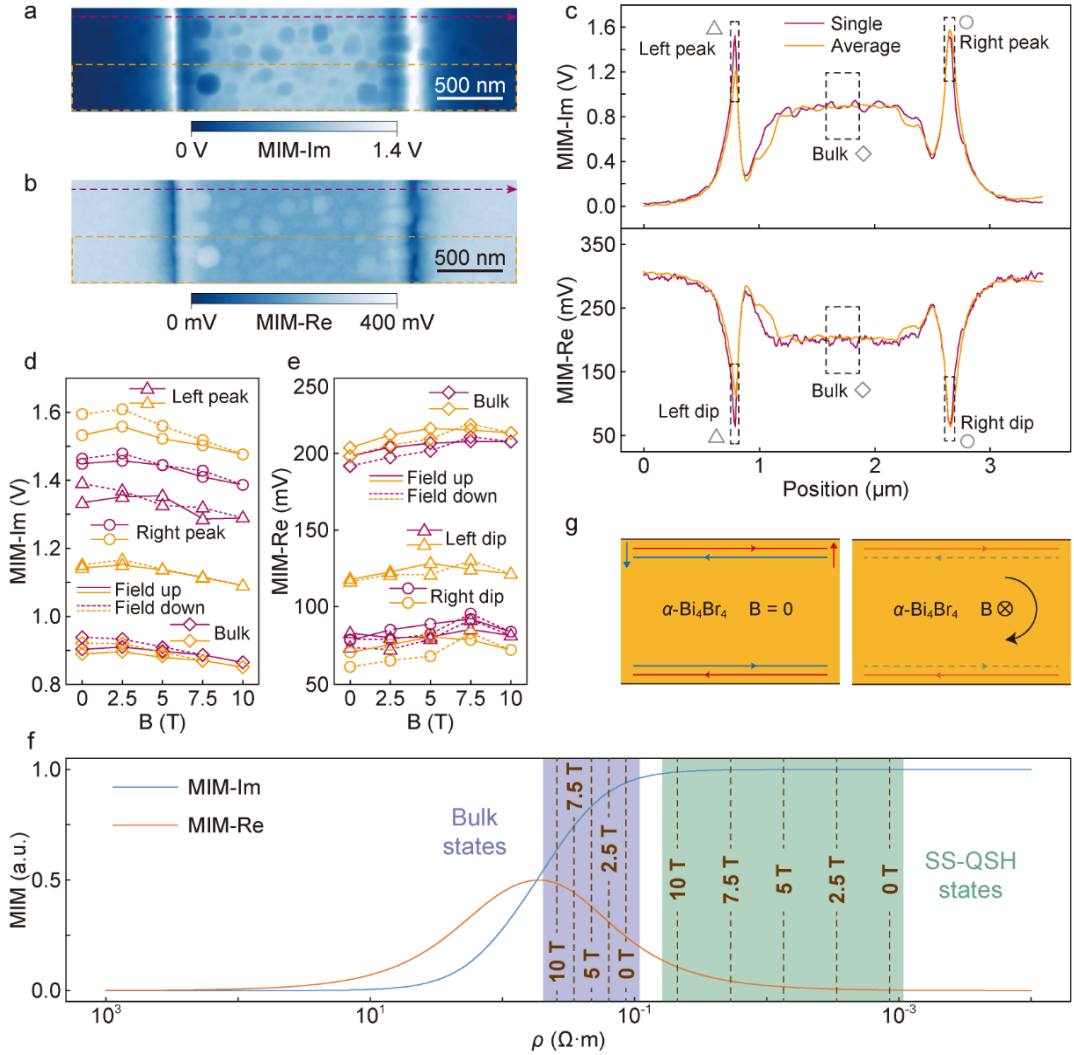

**Figure 4.** Magnetic-field dependence of SS-QSH states. (a, b) MIM-Im and MIM-Re images of a nanowire measured at 4.2 K, 0 T. (c) MIM-Im and MIM-Re line profiles from (a) and (b). Single-profiles are obtained along the purple dashed arrows and Average-profiles are obtained by averaging the yellow dashed rectangle area. (d, e) Magnetic-field response of SS-QSH and bulk states. The values of peaks and dips are extracted by averaging their corresponding gray dashed rectangles in (c). Single- and Average-profile data are marked with purple and yellow, respectively. Continuous field-up and field-down (solid and dashed lines, respectively) measurements are conducted. (f) Estimated resistivity of SS-QSH and bulk states in simulated response curves. The light green and purple areas represent their possible ranges and are intentionally enlarged for clarity. (g) Schematics illustrating magnetic-field effects on SS-QSH states. Left and right panels: Counterpropagating helical QSH states (spin-up in red, spin-down in blue) with and without vertical magnetic field.

**Magnetic-field dependence of SS-QSH and bulk states**

We also examined the magnetic-field dependence of SS-QSH and bulk states. MIM-Im and MIM-Re images of another $\alpha$-Bi$_4$Br$_4$ nanowire were simultaneously acquired at 4.2 K under zero magnetic field (Figs 4a and 4b). Complete magnetic-field data are presented in Fig. S9. The shapes of line profiles (Fig. 4c) resemble those observed in the temperature-dependent measurements (Fig. 3c), except that the bulk signals in Re-channel appear lower than those of the substrate SiO$_2$. This discrepancy is attributed to differences in resistivity between different nanowires. SS-QSH and bulk states exhibit similar behaviors in response to the applied vertical magnetic field (Figs 4d and 4e). With increasing magnetic field, MIM-Im signals decrease monotonically while MIM-Re signals increase monotonically, indicating that the conductivities of both SS-QSH and bulk states are suppressed (see illustration in the colored region of Fig. 4f). Compared with those in Fig. 3f, SS-QSH states remain in the highly conductive region of the simulated response curves, whereas bulk states are still located in the intermediate region, but shifted toward the conductive direction (Fig. 4f).

Despite similar trends under increasing magnetic fields, the underlying suppression mechanisms of SS-QSH and bulk states are distinct. For bulk states, the reduction in conductivity originates from magnetoresistance induced by high fields, a phenomenon that has been extensively reported [31,34,37]. In contrast, for SS-QSH states, the suppression arises from two factors: first, time-reversal-symmetry breaking opens a gap in SS-QSH states and leads to an overall weakening; second, the magnetic field induces an imbalance between the two counterpropagating spin channels, preferentially suppressing one channel (for example, spin-down channel in Fig. 4g) over the other (spin-up channel in Fig. 4g).

**Evidence for stair-stepped model**

The temperature and magnetic-field responses confirm the topological character of the SS-QSH states. To further clarify their stair-stepped origin, we measured a nanowire with continuously varying height (Figs 5a and 5b). Figure 5c shows profiles of the height, left/right peaks, and bulk signals aligned parallel to the nanowire (along the dashed arrows in Figs 5a and 5b). The signal strength of the left/right peaks clearly decreases as the nanowire height is reduced (left and right peak panels in Fig. 5c). According to the MIM detection mechanism (Fig. S2), this attenuation stems not from a direct drop in conductivity, but from the reduced number of Bi$_4$Br$_4$ layers, and hence of available monolayer-step QSH states, as illustrated schematically in Fig. 5d. The conductivity of each decoupled monolayer-step state is expected to remain independent of total height.

Additional support for the SS-QSH picture comes from the bulk-signal behavior (bulk panel in Fig. 5c). The region from the onset of height reduction to the nanowire tail can be viewed as a gradually descending special "side facet" (Figs 5c and 5d). Compared with constant-height bulk regions (delimited by the left gray dashed line in Fig. 5c), this zone shows enhanced bulk signals because it contains sparsely distributed conducting SS-QSH states superimposed on the bulk response. These SS-QSH states are not directly resolved in the MIM-Im image (Fig. 5b) because the signals from an individual step are weaker than that of the semiconducting bulk—especially at 77 K, where the bulk Fermi level remains in the conduction band [33]. The steady rise in bulk signals toward the tail reflects the increasing slope of the height reduction, which corresponds to a higher density of monolayer-step states (height panel in Fig. 5c).

We also conducted simulations to reproduce our results (Fig. 5e, see more details in Figs S3d and S3e). By constructing a model rigorously based on the line profile of topography, FEA successfully reproduces similar profile of MIM signals. Moreover, the best fitting results indicate that $\rho(\alpha\text{-Bi}_4\text{Br}_4) = 0.735$ $\Omega\cdot$m and $\rho(\text{SS-QSH}) = 10^{-3}$ $\Omega\cdot$m, which is consistent with the estimated resistivity of SS-QSH and bulk states in Figs 3f and 4f, further confirming the validity of above explanations. In addition, we thoroughly discuss higher-order hinge states and exclude them as possible origin of our observations (Section 10 in Supplementary Material).

**SUMMARY**

In summary, we directly visualized the room-temperature SS-QSH states in $\alpha$-Bi$_4$Br$_4$ nanowires by MIM. They resided on side facets formed with multiple vertically stacked monolayer steps and were compatible with large-scale systems. We systematically investigated their topological robustness through temperature and magnetic measurements. By studying bulk and SS-QSH states in nanowires with varying height, we undoubtedly demonstrated their origin from stair-stepped model. We also successfully reproduced our results by FEA simulations, further supporting our conclusions. Based on the room-temperature stability and large-sized-system compatibility of the SS-QSH states, $\alpha$-Bi$_4$Br$_4$ exhibits considerable practical potential. Further experiments, such as studies of magnetic-field orientation and microwave frequency dependence [50], can be pursued. Moreover, the stair-stepped stacking model demonstrated here is readily extendable to other high-temperature QSH materials, broadening the scope for applications. Future efforts may capitalize on the robustness of SS-QSH states to design hybrid topological circuits. Integration with superconductors or magnets could induce proximity effects and access advanced quantum phases. Van der Waals heterostructures with other 2D materials may enable gate-tunable topological devices. Developing scalable nanofabrication protocols would allow the realization of interconnected networks of topological channels, advancing large-scale quantum electronics. The high-temperature operation of this material platform also supports its use for hosting Majorana zero modes, thereby accelerating the development of energy-efficient topological circuits under ambient conditions.

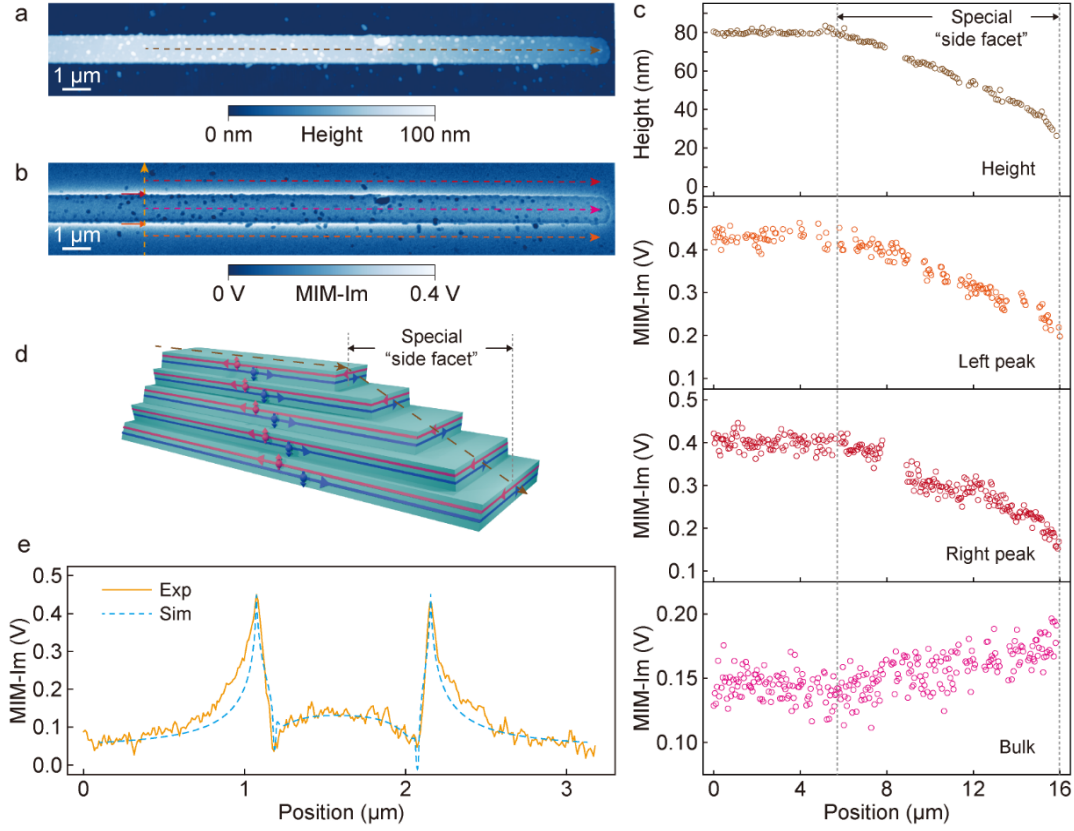

**Figure 5.** Dependence of SS-QSH states on nanowire height. (a, b) AFM topography and MIM-Im images of a nanowire with continuously varying height measured at 77 K. (c) Height (brown), left/right-peaks (orange/red) and bulk (pink) profiles along color-matched dashed arrows in (a) and (b). Profile positions of QSH-related peaks are slightly offset to avoid overlap. A special 'side facet' is identified between two gray dashed lines. Defect-affected data are excluded. (d) Schematic of a nanowire with continuously varying height. Monolayer-step QSH states are illustrated within each $Bi_4Br_4$ layer. Brown dashed arrows mark the same height-profile paths in (a) and the special 'side facet' is also labeled as that in (c). (e) Fitting of experiemntal profile along yellow dashed arrow in (b) with simulated profile via FEA.

## FUNDING
This work was supported by the National Natural Science Foundation (Grants No. 92365201, No. 52388201, No. 11427903, No. 12321004), the National Key Research and Development Program of China (Grant Nos. 2022YFA1403100, 2020YFA0308800, 2022YFA1403400), the Innovation Program for Quantum Science and Technology (2021ZD0302402), the Beijing Natural Science Foundation (Grant No. Z210006), and the Beijing National Laboratory for Condensed Matter Physics (Grant No. 2023BNLCMPKF007).

## AUTHOR CONTRIBUTIONS
We thank X. Wan for helpful discussions. W.L. and Q.-K.X. conceived and supervised the research project. Z.H., Y.W. and K.X. performed the MIM experiments. Y.Z., Z.W. and Y.Y.

provided the samples. Y.Z., X.L. and H.Y. performed the structure characterizations of nanowires. W.L., Z.H., Y.W. and Y.Y. analyzed the data. W.L. and Z.H. wrote the manuscript with input from all other authors.

**Competing financial interests**
The authors declare no competing financial interests.